\documentstyle[epsf,epsfig,psfig,shadow]{AGILEpsr}

\def\lesssim{\,\hbox{\lower0.6ex\hbox{$\sim$}\llap{\raise0.6ex\hbox{$<$}}}\,}
\def\gtrsim{\,\hbox{\lower0.6ex\hbox{$\sim$}\llap{\raise0.6ex\hbox{$>$}}}\,}

\begin{opening}

\title{A Hard X--ray View of Accreting X--ray Binary Pulsars}
\author{Mauro Orlandini}
\institute{Istituto di Astrofisica Spaziale e Fisica Cosmica 
  (IASF/CNR) --- Sezione di Bologna}
\date{} 
\end{opening}

\begin{document}

\oddpagefooter{}{}{} 
\evenpagefooter{}{}{} 
\medskip  

\begin{abstract}
The study of the hard ($E\gtrsim 10$~keV) energy spectra of X--ray
binary pulsars can give a wealth of information on the physical
processes that occur close to the neutron star surface. Extreme matter
regimes are probed, and precious information on how matter and radiation
behave and interact in critical conditions can be obtained. We will
give an overview on the most recent results obtained by RXTE and
BeppoSAX on this class of objects, in order to pass the baton onto just
launched experiments, like INTEGRAL, or soon to be launched, like AGILE
and ASTRO-E2.
\end{abstract}

\medskip

\section{Introduction}

The discovery of X--ray emission from celestial objects further extended
the knowledge of our Universe, which resulted very different from the
quiet and calm Universe the astronomers of the last centuries described. 
Indeed, it was just at the beginning of the 1960's that, with the advent
of stratospheric balloons and rockets, it was possible to launch outside
our atmosphere some Geiger counters (Giacconi et~al.\ 1962). In this way,
the first discrete X--ray source in our Galaxy was discovered: Sco X--1.
In 1966 the first optical counterpart to a galactic X--ray source, Sco
X--1, was identified with an old 12$^{\rm th}$--13$^{\rm th}$ magnitude
star (Sandage et~al.\ 1966). In the following years a theoretical model
was developed according to which galactic X--ray sources are close,
interacting binary systems composed by a ``normal'' star and a compact
object (Shklovskii 1967). In the same years it was understood that
spherical accretion could become non symmetric, leading to the formation
of an accretion disk around the compact object, if the accreting matter
possesses enough angular momentum (Prendergast \& Burbidge 1968). 

But the decisive step toward the comprehension of this class of objects
was achieved with the discovery of pulsed emission from some X--ray
sources (Giacconi et~al.\ 1971; Tananbaum et~al.\ 1972).  Indeed, the
variability on short time scale --- for example the first discovered
X--ray binary pulsar, Cen X--3, spins at about 4.8~s --- implies a small
emitting region. Furthermore, because the system is not destroyed by the
centrifugal force it is necessary that at the surface of the emitting
object the gravitational force is greater than the centrifugal one. This
implies $\Omega_p \gtrsim \sqrt{G\langle\rho\rangle}$, where $\Omega_p$ is
the spin frequency, $G$ the gravitational constant and
$\langle\rho\rangle$ the object mean density. The observed values of
$\Omega_p$ imply $\langle\rho\rangle \gtrsim 10^6$ g/cm$^3$, and therefore
the compact nature of the object responsible of the pulsed X--ray emission
was established.  The only compact object able to explain all the
phenomena observed in X--ray pulsars, as pulse period range and surface
magnetic field strength, is a neutron star (see, e.g., Shapiro \&
Teukolsky 1983). 

The standard model explains the X--ray emission as due to the conversion
of the kinetic energy of the in-falling matter (coming from the intense
stellar wind of an early optical star, in this case we speak about
wind-fed binaries, or from an accretion disk due to Roche-lobe overflow,
and this is the case of disk-fed binaries) into radiation, because of the
interactions with the strong magnetic field of the neutron star, of the
order of $10^{11}$--$10^{13}$~G\footnote{Obtained from conservation of
magnetic flux during the process of collapse from a ``normal'' star
($B\sim 10-100$~G, $R\sim 10^6$~Km) to a neutron star ($R\sim 10$~Km)
and lately confirmed by the observation of cyclotron resonance features
(see below).}. 
The dipolar magnetic field of the compact object drives the accreted
matter onto the magnetic polar caps, and if the magnetic field axis is not
aligned with the spin axis then the compact object acts as a
``lighthouse'', giving rise to pulsed emission when the beam (or the
beams, according to the geometry) crosses our line of sight. 

Subsequent observations clearly demonstrated the binary nature of these
objects by observations of X--ray eclipses and Doppler delays in the pulse
arrival times (Schreirer et~al.\ 1972). From these measurements it was
possible to ``solve'' the binary systems, obtaining masses in agreement
with that expected for a neutron star.
But for some X--ray pulsars, with pulse periods in the 5--12~s range, every
attempt to find signatures of binary motion was unsuccessful. It was
later recognized that they form a class of their own, the so-called
anomalous X--ray pulsars, in which their X--ray emission is due to the
conversion of the magnetic field energy into radiation (Thompson \&
Duncan 1995; 1996).  We will not discuss here on this quite
interesting class (see, e.g., Mereghetti et~al.\ 2002 for a review). 

\section{Astrophysics of X--ray binary pulsars}

\begin{figure}
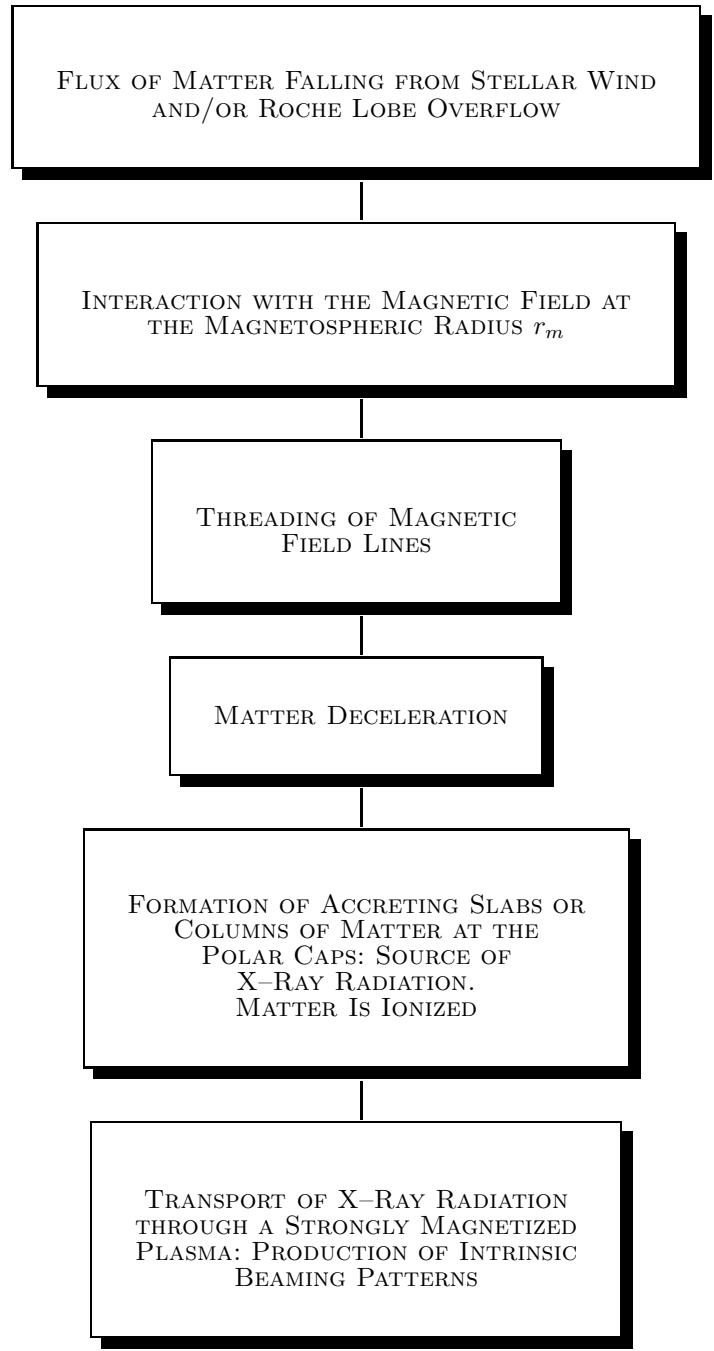

\begin{center}
\shabox{\rule[-0.5cm]{0cm}{1.8cm}\shortstack{\sc ~~~Flux of Matter
Falling from Stellar Wind~~~ \\ \sc and/or Roche Lobe Overflow}}

\rule{0.2mm}{0.5cm} 

\shabox{\rule[-0.5cm]{0cm}{1.8cm}\shortstack{\sc ~~~Interaction with the
Magnetic Field at~~~ \\ \sc the Magnetospheric Radius $r_m$}}

\rule{0.2mm}{0.5cm}

\shabox{\rule[-0.5cm]{0cm}{1.8cm}\shortstack{\sc ~~~Threading of
Magnetic~~~ \\ \sc Field Lines}}

\rule{0.2mm}{0.5cm}

\shabox{\rule[-0.5cm]{0cm}{1.2cm}\shortstack{\sc ~~~Matter Deceleration~~~}}

\rule{0.2mm}{0.5cm}

\shabox{\rule[-0.5cm]{0cm}{2.8cm}\shortstack{\sc ~~~Formation of Accreting
Slabs or~~~ \\ \sc Columns of Matter at the \\ \sc Polar Caps: Source
of \\ \sc X--Ray Radiation. \\ \sc Matter Is Ionized}}

\rule{0.2mm}{0.5cm}

\shabox{\rule[-0.5cm]{0cm}{2.5cm}\shortstack{\sc Transport of X--Ray
Radiation \\ \sc ~~~through a Strongly Magnetized~~~ \\ \sc Plasma:
Production of Intrinsic \\ \sc Beaming Patterns}}
\end{center}
\caption[]{Block diagram of the physical processes of production and
emission of the X--ray flux in a X--ray binary pulsar.}
\label{blockdiagram}
\end{figure}

The main astrophysical problem connected with the physics of X--ray
pulsars is that we cannot use a linearized theory but we are forced to use
the full magneto-hydrodynamical one. This is due to the fact that the
coupling constants of the interactions (in this case gravitational and
magnetic)  are so large that a series expansion is not possible.
Furthermore, the highly non-linear nature of the problem makes its
treatment very difficult. In Fig.~\ref{blockdiagram} a sort of block
diagram of the physical processes of production and emission of the X--ray
flux in a X--ray binary pulsar is shown.

%
%

Each block in Fig.~\ref{blockdiagram} is characterized by its typical
physical processes and characteristic time and length scales.  In the
first block we deal with the problem of determining how much matter is
captured by the neutron star gravitational field and how its angular momentum
is transferred to the neutron star (this will have important consequences on
the spinning behavior of the neutron star). All the matter swept by the
neutron star inside a distance called ``accretion radius'' will be
captured and accreted. This radius depends on the relative velocity of the
wind matter with respect to the neutron star (Bondi \& Hoyle
1944):

\begin{equation}
r_a = \frac{2GM_x}{v^2_{\rm rel} + c^2_s} \approx
      \frac{2GM_x}{v^2_{\rm orb} + v^2_{\rm win}}
\end{equation}

\noindent where $M_x$ is the neutron star mass, $c_s$ is the sound speed
(negligible because the wind matter is supersonic; Elsner \& Lamb
1977), and $v_{\rm orb}$ and $v_{\rm win}$ are the orbital and wind
velocity, respectively. We expect that the characteristic time scales in
this ``block'' be dynamical, of the order of 100--1000~s. 

It is worthy to introduce here another scale length, connected to the
rotation of the neutron star. In order for matter to be accreted it is
necessary that the neutron star does not rotate so fast that plasma is
expelled because of the centrifugal force. The distance at which there is
balance between these two forces is called corotation radius, defined as

\begin{equation}
r_c = \left(\frac{GM_x}{\Omega_p^2}\right)^{1/3} =
1.5\times 10^8 \; P^{2/3} m^{1/3}\ {\rm cm}
\end{equation}

\noindent where $P$ is the pulse period in seconds, and $m$ the neutron
star mass in solar units. 

At some distance from the neutron star surface, called magnetospheric
radius,
the magnetic field of the neutron star becomes the main interaction which
drives the motion of the captured matter toward the stellar surface. At
this distance matter is halted by the very
strong magnetic field of the neutron star and accretion can occur only if
matter can penetrate the shock layer by means of magneto-hydrodynamical
instabilities (Elsner \& Lamb 1977). From its definition, the
magnetospheric radius will depend on the magnetic field strength and the
ram pressure of the accreted matter (see, e.g., Frank, King \& Raine 1985): 

\begin{equation}
r_m = \left\{
\begin{array}{l}
5.1\times 10^8\; \xi^{2/7} \mu_{30}^{4/7} m^{-1/7} \dot{M_{16}}^{-2/7}\
{\rm cm} \\ \\
2.9\times 10^8\; \xi^{2/7} \mu_{30}^{4/7} m^{-1/7} r_6^{-2/7}
\epsilon_{0.1}^{2/7} L_{37}^{-2/7}\ {\rm cm}
 \end{array}  \right.
\end{equation}

\noindent where $\xi\lesssim 1$, $\mu_{30}$ is the dipolar magnetic moment
in units of $10^{30}$ G~cm$^3$,
$\dot{M}_{16}$ the mass accretion rate onto the neutron star in units of
$10^{16}$ g/s, $r_6$ the neutron star radius in units of $10^6$ cm,
$\epsilon_{0.1}$ the accretion efficiency in units of 0.1, and $L_{37}$
the X--ray luminosity in units of $10^{37}$ erg/s.  Because the main
physical processes occurring in this ``block'' are magneto-hydrodynamical
instabilities, the characteristic time scales will be 0.1--10~s.

According to fastness of rotation, different kind of instabilities will
determine plasma penetration (Arons \& Lea 1976): in the case of slow
rotators\footnote{Slow rotators have pulse frequency so small that
rotation can be neglected in all the equations describing the physics of
accretion. A quantitative parameter which measures the importance of
rotation is the ``fastness parameter'', defined as (Elsner \& Lamb 1977):
$\omega_k \equiv \Omega_p/\Omega_k(r_m)$, where $\Omega_k$ is the angular
velocity of matter orbiting into Keplerian orbits. If $\omega_k \ll 1$
rotation can be neglected.} plasma penetration will occur mainly by means
of gravity-driven interchange (Rayleigh-Taylor) instability.  Decreasing
the pulse period, the shear between the plasma and the magnetosphere
becomes more and more important, leading to the stabilization of the
magnetopause with respect to Rayleigh-Taylor instability, and giving
rise to the onset of Kelvin-Helmholtz instability.

By comparing the two length scales, magnetospheric and corotation radii,
it is possible to distinguish two accretion regimes: if $r_c\gtrsim r_m$
then at the magnetospheric limit, when plasma penetrates the
magnetosphere, the centrifugal force is smaller than the magnetic force,
and therefore matter can be accreted.  This is the so-called {\em
accretor} regime.  On the other hand, if $r_c\lesssim r_m$ then the
centrifugal force inhibits matter from being channeled and is swept away.
This is the so-called {\em propeller} regime (Illarionov \& Sunyaev 1975). 

Once matter has penetrated the magnetosphere, it will follow the magnetic
field lines up to the magnetic polar caps of the neutron star, where it
will be decelerated. If the amount of matter falling on the polar caps is
high enough that a X--ray luminosity greater than about $10^{37}$ erg/s is
reached, then a radiative shock will form (Basko \& Sunyaev 1976). In this
case an accretion column just above the polar cap will form; this
accretion column will be optically thick to X--rays, therefore radiation
will be emitted mainly {\em sideways}.  Radiation is emitted mainly in a
direction perpendicular to the magnetic field lines and we call this
pattern ``fan beam emission pattern''. 

On the other hand, if the X--ray luminosity is lower than about $10^{37}$
erg/s, then the radiative shock will not form and matter will be able to
reach the neutron star surface. In this case we will have the formation of
an emitting ``slab'' and radiation will be emitted mainly in a direction
parallel to the magnetic filed lines. We call this pattern ``pencil beam
emission pattern''. Because the main physical processes occurring in this
``box'' are Compton heating and cooling, bremsstrahlung and Coulomb
interactions, the characteristic time scale will be $\lesssim 0.001$~s. 

Before leaving the neutron star, radiation interacts with the surrounding
medium and the very strong magnetic field. As briefly discussed above,
this interaction is very difficult to treat because of the impossibility
to linearize the theory due to our substantial ignorance on very
strong magnetic fields (see Harding 2003 for a recent review). One of
the most important consequences of the presence of a strong magnetic
field is the quantization of the electron motion in the direction
transverse to $B$: this leads to the so-called Landau levels (see
M{\'e}sz{\'a}ros 1992 for a complete treatment of this topic). In the
non-relativistic case, the energy associated to each level is given by

\begin{equation}
\hbar\omega_n = n\,\hbar\omega_c
\label{omegan}
\end{equation}

\noindent where the Larmor gyro-frequency $\omega_c$ is defined as
$eB/\gamma m c$, with $\gamma$ the Lorentz factor, $e$ and $mc$ the
electron charge and momentum, respectively. As an aside, from
Eq.~\ref{omegan} we have that $E_n = 11.6\cdot B_{12}$~keV, where $B_{12}$
is the magnetic field strength in units of $10^{12}$~G, and therefore
observable in the hard X--ray energy band.  When relativistic
corrections are taken into account a slight non-harmonicity is introduced in
the Landau levels. Indeed, we have

\begin{equation}
\hbar\omega_n = mc^2\,
  \frac{\sqrt{mc^2 + 2n\hbar\omega_c\,\sin^2\theta} - 1}{\sin^2\theta}
\end{equation}

\noindent where $\theta$ is the angle between the line of sight and $B$.

Due to the existence of these levels, an electromagnetic wave which
propagates in such a plasma will have well defined polarization normal
modes, {\it i.e.\/} the medium will be birefringent (Ginzburg 1970). 
Furthermore, for magnetic fields not far from the critical value of
$1.414\cdot 10^{13}$~G, an important r\^ole is played by virtual
electron-positron pairs. The corresponding virtual photons dominate the
polarization
properties of the medium and therefore the radiative opacity of the
plasma. This means that the scattering cross sections of
X--rays are strongly anisotropic and energy dependent (Herold 1979).

It is important to stress that the cyclotron {\em absorption} cross
section is resonant for energies equal to the gyro-magnetic (Larmor) 
frequency $\omega_c$.  Once the electron absorbs a photon it (almost) 
immediately de-excites on a time scale $t_r\sim 2.6\times
10^{-16}\,B_{12}^{-1}$~sec (M{\'e}sz{\'a}ros 1992).  This has important
consequences for the {\em scattering} cross sections. Indeed, while a
scattering process involves two photons (one going in, one going out),
absorption (or emission) processes involve only one photon. Therefore one
expects that the two cross sections are different. This is not true just
because an absorbed photon is immediately re-emitted, and therefore the
absorption-emission process is equivalent to a scattering. Therefore
photons with frequency close to $\omega_c$ will be scattered out of the
line of sight, creating a drop in their number. Cyclotron ``lines''
observed in the spectra of X--ray binary pulsars are therefore {\em not}
due to absorption processes, but are due to scattering of photons resonant
with the magnetospheric electrons (as it occurs for the Fraunhofer lines
in the Solar spectrum).  This is why we will not use the term cyclotron
lines but the more appropriate ``cyclotron resonant features'' (CRFs).

\section{Observation of X--ray binary pulsars}

As we pointed out in the previous section, a great deal of information can
be obtained by the observation of the hard ($E\gtrsim 10$~keV) spectra of
X--ray binary pulsars.  The advantage of focusing on the hard X--rays is
that in this energy range we are observing phenomena that occur close to
the neutron star surface and that are less subject to absorption phenomena
that alter the emergent spectra. An overview of past hard X--ray
observations is already available in literature (see e.g.\ Orlandini \&
Dal~Fiume 2001) therefore we will focus on recent results that can be used
as starting point for present and future missions, like INTEGRAL, AGILE and
ASTRO-E2. 

The best recent X--ray telescopes suited for the study of X--ray binaries 
are (or have been) RXTE and BeppoSAX (in particular the two high energy
instruments HEXTE and PDS). The advantage of BeppoSAX with respect to its
US fellow was its larger band-pass, fundamental for the reconstruction
of the continuum, and the lower intrinsic background that allowed a better
sensitivity. Anyway, both satellites gave a wealth of new information
and opened a new era for the study of X--ray pulsars, passing the baton onto
the just launched INTEGRAL. 

In Table~\ref{tablepuls} we list the X--ray binary pulsars observed by
BeppoSAX and about which we will discuss in this paper (for a complete
discussion on the RXTE observations of X--ray pulsars see Coburn et~al.\
2002). For each of them we present the value of its CRF energy, if
observed. 

\begin{table}
\caption{BeppoSAX observations of X--ray binary pulsars}
\label{tablepuls}
\begin{center}
\def\arraystretch{1.7}
\begin{tabular}{|l|l|c|c|l|}
\hline\hline\noalign{\smallskip}
\multicolumn{1}{|c}{Source}& 
\multicolumn{1}{|c}{Obs Date}  & 
\multicolumn{1}{|c}{E$_{\rm cyc}$ (keV)} &
\multicolumn{1}{|c}{FWHM (keV)} & 
\multicolumn{1}{|c|}{References} \\
\hline\hline\noalign{\smallskip}
{\bf 4U0115+63 (M)}    & 20 Mar 1999 & $12.78\pm 0.08$ & $3.58\pm 0.33$ & Santangelo et~al.\ 1999 \\
{\bf 4U1538--52 (M)}   & 29 Jul 1998 & $21.5\pm 0.4$   & $6.7\pm 1.2$   & Robba et~al.\ 2001 \\
\underline{\bf Cen X--3 (M?)}& 27 Feb 1997 & $28.5\pm 0.5$   & $7.3\pm 1.9$   & Santangelo et~al.\ 1998 \\
\underline{\bf XTE J1946+27} & 09 Oct 1998 & $33\pm 4$       & $16\pm 2$      & Orlandini et~al.\ 2001 \\
\underline{\bf OAO1657--415} & 04 Sep 1998 & $36\pm 2$       & 10             & Orlandini et~al.\ 1999 \\
\underline{\bf 4U1626--67}  & 06 Aug 1996 & $38.0\pm 0.9$   & $11.8\pm 1.7$  & Orlandini et~al.\ 1998b \\
{\bf 4U1907+09 (M)}    & 29 Sep 1997 & $38.3\pm 0.7$   & $9.7\pm 2.3$   & Cusumano et~al.\ 1998 \\
{\bf Her X--1}         & 27 Jul 1996 & $42.1\pm 0.3$   & $14.7\pm 1.1$  & Dal~Fiume et~al.\ 1998 \\
\underline{\bf GX301--2} & 24 Jan 1998 & $49.5\pm 1.0$ & $17.9\pm 2.5$  & Orlandini et~al.\ 2000 \\
\underline{\bf Vela X--1}& 14 Jul 1996 & $54.8\pm 0.9$ & $25.0\pm 2.1$  & Orlandini et~al.\ 1998a \\
{\bf A0535+26}         & 04 Sep 2000 & $118\pm 20$     & $81\pm 50$     & Orlandini et~al.\ 2004 \\
{\bf GX1+4}            & 25 Mar 1997 & \ldots          & \ldots         & Israel et~al.\ 1998 \\
{\bf GS1843+00}        & 04 Apr 1997 & \ldots          & \ldots         & Piraino et~al.\ 2000 \\
{\bf X Persei}         & 09 Sep 1996 & \ldots          & \ldots         & Di~Salvo et~al.\ 1998 \\
\hline\noalign{\smallskip}
\multicolumn{5}{l}{Sources \underline{underlined} are discoveries made
by BeppoSAX --- M stands for multiple lines detected/suspected}
\end{tabular}
\end{center}
\end{table}

\subsection{The X--ray continuum}

The characterization of the continuum is of paramount importance for the
determination of the physical processes that are at play. As it should be
clear from the Introduction, the main physical process responsible for the
continuum emission in X--ray binary pulsars is Compton scattering. 
Broadly speaking, there are two regimes as a function of the
comptonization parameter $y$, that give rise to two completely different
emergent spectra. If $y\ll 1$ only coherent scattering will be important,
and the emergent spectrum will be a blackbody spectrum or a ``modified''
blackbody spectrum according whether the photon frequency is lower or
greater than the frequency at which scattering and absorption coefficients
are equal (Rybicki \& Lightman 1975). 

On the other hand, if $y\gg 1$ then inverse Compton scattering can be
important. If we define a frequency $\omega_{co}$ such that
$y(\omega_{co})=1$, then for $\omega\gg\omega_{co}$ the inverse Compton
scattering is saturated and the emergent spectrum will show a Wien hump,
due to low-energy photons up-scattered up to $\hbar\omega\sim 3kT$
(Rybicki \& Lightman 1975). In the case in which there is not saturation a
detailed analysis of the Kompaneets equation shows that the spectrum will
have the form of a power law modified by a high energy cutoff (Rybicki \&
Lightman 1975;  Sunyaev \& Titarchuk 1980). 

On the observational point of view, the first attempt to describe X--ray
pulsar spectra was done by White et~al.\ (1983) who introduced a cutoff
power law that mimicked the unsaturated inverse Compton process. In order
to ``smooth'' the break around the cutoff energy, Tanaka (1986) introduced
the so-called Fermi-Dirac cutoff, but there was no physical meaning for the
fitting parameters.  The discovery of a correlation between the cutoff
energy and the CRF energy by Makishima and Mihara (1992) led Mihara (1995)
to introduce the first analytical fitting law with a clear physical
meaning of its parameters, the so-called NPEX (Negative Positive
Exponential) model: 

\begin{equation}
{\rm NPEX}(E) = (AE^{-\alpha} + BE^{+\beta})
  \exp\left(-\frac{E}{kT}\right) \quad.
\label{npex}
\end{equation}

This model is quite successful in describing the X--ray pulsar spectra
observed by Ginga in the 3--30~keV. Its components have also a physical
meaning, because it mimics the saturated inverse Compton spectrum. 
Furthermore, because the (non relativistic) energy variation of a photon
during Compton scattering is $\Delta E/E = (4kT-E)/mc^2$ (Rybicki \&
Lightman 1975)  then when $E=E_c$ the medium is optically thick and
therefore $E_c\sim 4kT$.

From an observational point of view, the X--ray pulsars observed by
BeppoSAX and listed in Table~\ref{tablepuls} cannot be well fit by
Eq.~\ref{npex}. In particular, we find that their continuum can be
described in terms of (i) a black-body component with temperature of few
hundreds eV; (ii) a power law of photon index $\sim$1 up to $\sim$10~keV;
and a (iii) a high energy ($\gtrsim 10$~keV) cutoff that makes the
spectrum rapidly drop above $\sim$40--50~keV. 

Particular care must be taken in the description of the cutoff. Indeed, an
incorrect parametrization of the change of slope can introduce features
that are not real but dependent on the choice of the functional adopted to
model the continuum. In particular, for the X--ray pulsar OAO1657--415 it
was clearly detected a two-step steepening of the spectrum (Orlandini
et~al.\ 1999): a first change of slope occurring in the $\sim$10--20~keV
range, while a second steepening occurring at higher energies. By using a
single-step steepening model will give as a result the creation of
features that could be erroneously attributed to CRFs.  We think this is
the case for the claimed CRF at $\sim$25~keV in Vela X--1 (Orlandini
et~al.\ 1998a; Kreykenbohm et~al.\ 2002).  Indeed, by using a smoother
description of the cutoff, La~Barbera et~al.\ (2003) showed that the
$\sim$25~keV CRF is not necessary for fitting the Vela X--1 spectra.

Another possible source of confusion could be raised by features due 
instrumental
effects.  A standard way to remove these effects is to normalize the
source observed spectrum, channel by channel, to the Crab spectrum.  The
Crab nebula is considered a ``standard candle'' in X--ray astronomy,
because of its brightness, steadiness, and featureless single power law
spectrum. If a feature is instrumental, then it should be washed out in
the Crab ratio. As we will show in the next section, we will use this tool
to clearly identify features in the spectra of X--ray pulsars.

\subsection{Cyclotron resonance features}

The very first observation of a CRF in a spectrum of an X--ray pulsar was
performed in 1978 when Tr{\"u}mper et~al.\ observed a $\sim$35~keV CRF in
the spectrum of Her X--1. A while later, in the spectrum of the transient
X--ray pulsar 4U0115+63 not only the fundamental but also the first harmonics
was observed (Wheaton et~al.\ 1979).  Observations of CRFs in
other X--ray pulsars showed that they are a quite common phenomenon in
this class of objects.  At the beginning they were described empirically as
an {\em additive} Gaussian in absorption ({\em i.e.\/} a Gaussian function
with negative normalization).  Because CRFs are broad features, this
modeling did not fit well because it depends on the adopted continuum,
therefore Soong et~al.\ (1990)  introduced the {\em multiplicative}
Gaussian in absorption, defined as

\begin{equation}
{\rm GAUABS}(E) =
\left[1-I\exp\left(-\frac{(E-E_c)^2}{2W^2}\right)\right] \quad.
\label{gausabs}
\end{equation}

Mihara (1995), on the other hand, introduced a different description of
the CRF in terms on a Lorenzian function, led by the fact that the
cyclotron scattering cross section has this form. The so-called cyclotron
absorption function has the form

\begin{equation}
{\rm CYAB}(E) = \exp \left(-\frac{\tau(WE/E_c)^2}{(E-E_c)^2 + W^2} \right)
\quad.
\label{cycabs}
\end{equation}

The BeppoSAX observations showed us that Eq.~\ref{gausabs} is a better
description of CRFs: the reason is that Eq.~\ref{cycabs} is deeply
connected with the NPEX continuum. Indeed, the CYAB description of the
CRF should be used {\em only} together with the NPEX continuum. We
observed that the inclusion of Eq.~\ref{cycabs} on a power law continuum
results in changing the power law parameters, too. This is the reason why
all the CRF energies listed in Table~\ref{tablepuls} were obtained from
Eq.~\ref{gausabs}.

\begin{figure}
\vspace{-160pt}
{\centering\epsfig{file=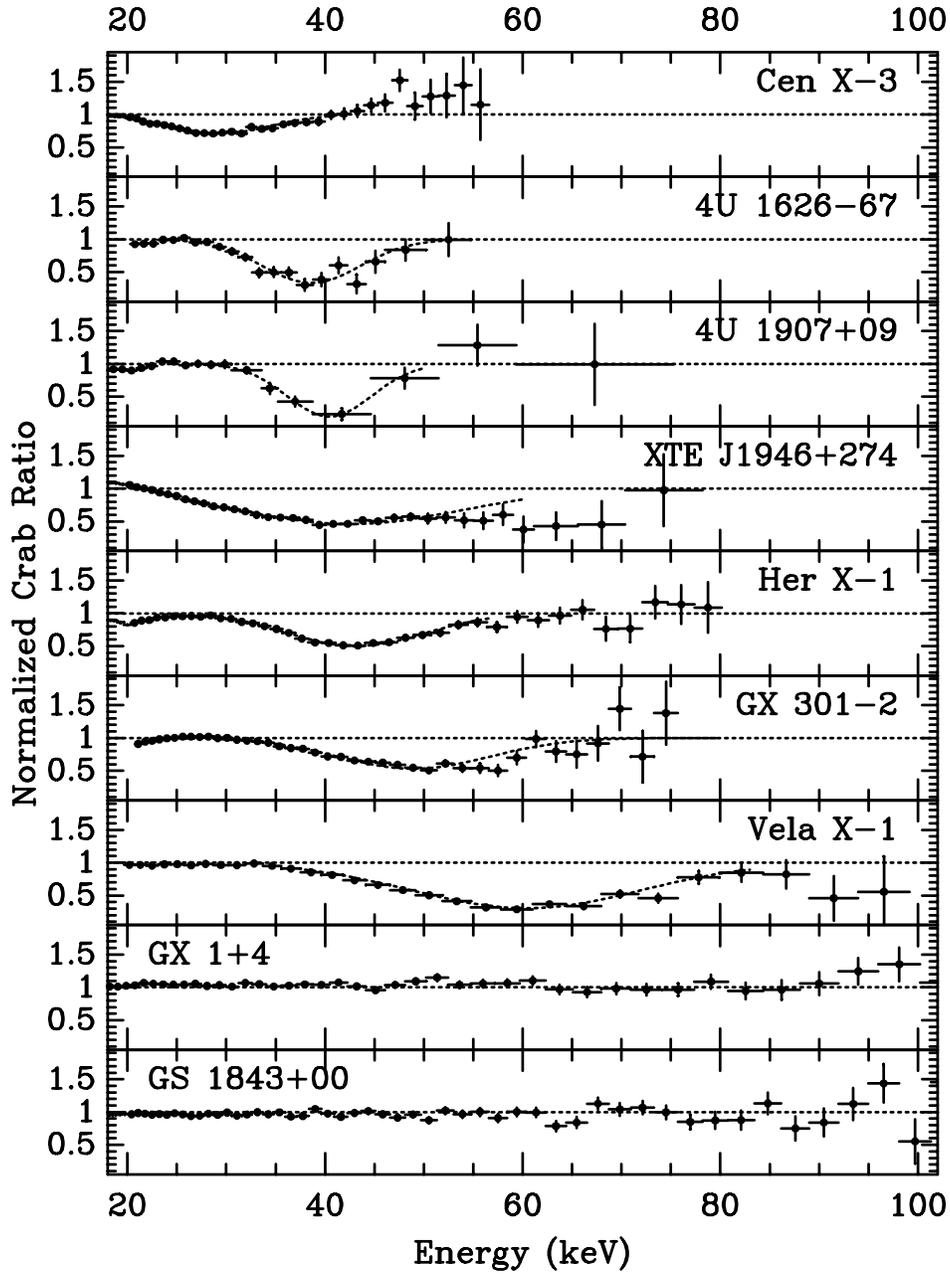,height=0.95\textheight}}
\vspace{1cm}
\caption[]{Normalized Crab Ratio computed on some of the X--ray pulsars
listed in Table~\ref{tablepuls}. Note that the higher the CRF energy, the wider
the feature. For GX1+4 and GS1843+00 no CRFs are present up
to 100~keV, while it is worth noting that the Vela X--1 NCR does not
show any CRF at $\sim$25~keV.}
\label{crabrat}
\end{figure}

In order to better characterize CRFs we added a further step to the Crab
ratio described in the previous section. Indeed, by multiplying the ratio
by a $E^{-2.1}$ power law (the functional form of the Crab nebula
spectrum), and dividing by the functional describing the continuum shape
of the source, any feature above the continuum will be greatly enhanced.
This procedure (that we call Normalized Crab Ratio --- NCR) has been
successfully applied to the X--ray binary pulsars listed in
Table~\ref{tablepuls} and the result is shown in Fig.~\ref{crabrat} for
some of them. 

From this figure it is evident that the higher the CRF energy,
the wider the feature. This is easily understood in terms of Doppler
broadening of the electrons responsible of the resonance, and holds for
all the sources displaying single CRFs (Orlandini \& Dal~Fiume 2001). In
other words, it seems that the temperature of the electrons responsible of
CRFs is the same for all X--ray binary pulsars, and is in the range
$\sim$15--30 keV. This energy range is somehow ``critical'', as pointed
out before (see also Coburn et~al.\ 2002 for RXTE results), because it is the
range in which the spectrum of X--ray pulsars shows a change of slope.  On
the other hand, the same relation found for the fundamental does not hold
for higher CRF harmonics: this means that the temperature of the electrons
responsible of higher CRF harmonics is different from that of the
electrons responsible of the fundamental CRF.
It is also worth noting that the Vela X--1 NCR does not show any CRF at
$\sim$25~keV.

\section{Conclusions}

The study of X--ray binary pulsars, especially in the hard X--ray band,
received a new momentum from the results by BeppoSAX and RXTE. Besides the
direct measurement of the neutron star magnetic field strength, the
observation of CRFs can give hints on the physical processes occurring in
extreme condition of temperature, density, gravity and magnetic field.  It
will be up to new missions (INTEGRAL, AGILE, ASTRO-E2) to find the answers
to the issues that were raised by their predecessors.

\end{document}